%% file: Manuscript.tex
\documentclass[pra,reprint,a4paper,superscriptaddress,showpacs,amsmath, floatfix,amssymb,aps]{revtex4-2}
\usepackage{graphicx}% Include figure files
\usepackage{subfigure}
\usepackage{dcolumn}% Align table columns on decimal point
\usepackage{bm}% bold math
\usepackage{hyperref}% add hypertext capabilities
\usepackage{indentfirst}
\usepackage{braket}
\usepackage{amsmath}
\usepackage{ulem}
\usepackage{color}
\usepackage{makecell}
\usepackage{booktabs}
\usepackage{footnote}
\usepackage[ruled,vlined]{algorithm2e}
\hypersetup{colorlinks=true, citecolor=blue, urlcolor=blue, linkcolor=blue}
\begin{document}

\title{Ground state properties of multi-component bosonic mixtures: a Gutzwiller mean-field study}
\author{Chenrong Liu}
\email[Corresponding authors: ]{crliu@wzu.edu.cn}
\affiliation{College of Mathematics and Physics, Wenzhou University, Zhejiang 325035, China}
\affiliation{State Key Laboratory of Surface Physics and Department of Physics, Fudan University, Shanghai 200433, China}
\author{Po Chen}
\affiliation{College of Mathematics and Physics, Wenzhou University, Zhejiang 325035, China}
\author{Linli He}
\email[Corresponding authors: ]{Linlihe@wzu.edu.cn}
\affiliation{College of Mathematics and Physics, Wenzhou University, Zhejiang 325035, China}
\author{Fangfang Xu}
\email[Corresponding authors: ]{Fangfangxu@wzu.edu.cn}
\affiliation{College of Mathematics and Physics, Wenzhou University, Zhejiang 325035, China}
\date{\today}
\begin{abstract}
Using the single-site Gutzwiller method, we theoretically study the ground state and the interspecies entanglement properties of interexchange symmetric multi-component (two- and three-) bosonic mixtures in an optical lattice, and the results are generalized to an $n$-component ($n=2,3,4,\cdots$) system. We compute the mean-field phase diagram, the interspecies entanglement entropy, and the ground state spectral decomposition. Three phases namely the $n$-component Superfluid state (nSF), the $n$-component Mott insulator state (nMI), and the Super-counter-fluid state (SCF)  are observed. Interestingly, we find that there are $n-1$ SCF lobes to separate every two neighboring nMI lobes in the phase diagram. More importantly, we derive the exact general expression of the interspecies entanglement entropy for the SCF phase. In addition, we also investigate the demixing effect of an n-component mixture and demonstrate that the mixing-demixing critical point is independent of n.
\end{abstract}
\pacs{}
% insert suggested keywords - APS authors do not need to do this
%\keywords{}
\maketitle

%%%%%%%%%%%%%
\section{Introduction} 
Recently, an amount of theoretical and experimental research has been devoted to understanding the new physical phenomena of bosonic mixtures. Many of these studies are focused on the two-component bosonic system which can be experimentally realized with ultracold atoms in optical lattices \cite{2CBMExp1,2CBMExp2,2CBMExp3} and described by a two-component Bose-Hubbard model \cite{2CBMHam}. In this two-component bosonic mixture, a rich phase diagram emerges, such as the paired superfluid (PSF), the super-counter-fluid (SCF), the molecular superfluid,  the non-integer insulator, the charge density wave, and the novel magnetic states \cite{2CBMPhase1,2CBMPhase2,2CBMPhase3,2CBMPhase4,2CBMPhase5,2CBMPhase6, 2CBMPhase7}. In the SCF (PSF) phase, the equal-current flows of two components in opposite (same) directions are superfluid but the single atomic superfluid current flow is impossible \cite{PhysRevLett.92.030403}. It means that the particle-hole (particle-particle) pairing between the two species is formed in the SCF (PSF) phase. These pairings can be treated as composite bosons and they are condensed in the SCF (PSF) phase while the individual bosons are in a localized state.  From the perspective of mean-field theory, the SCF (PSF) phase is such a state in which the atomic superfluid order parameter $\langle a \rangle$ and $\langle b \rangle$ should be zero, but the composite particle superfluid order parameter $\langle ab^\dag\rangle$ ($\langle ab\rangle$) must be non-zero. Here, $a$ and $b$ label the bosonic annihilation operator of $a$ and $b$ species respectively.

If a third component is introduced, some exotic quantum phases which can not be observed in the two-component mixtures, become available. For instance, the formation of the Borromean droplet, where only the ternary bosons can form a self-bound droplet while any binary subsystems cannot, is identified \cite{3CBMPhase1}. Another example is that the polaron physical phenomenon can be induced in a highly imbalanced three-component bosonic mixture (One or two bosonic impurities are treated as the distinguished boson components) \cite{PhysRevA.104.L031301, Mistakidis_2022,theel2023crossover}. The phase diagrams of a three-component bosonic mixture in optical lattices are also investigated both with \cite{ZHANG2022128035, Zan} and without an external magnetic field \cite{3CBMPhase2}. However, these studies neglect the presence of pairing phases, such as the SCF phase, that are expected to exist in a bosonic mixture. Therefore, one of our goals in this paper is to determine the SCF region in the phase diagram for this three-component mixture. For simplicity, we assume an interexchange symmetry here, meaning each species is equivalent. 

On the other hand, the entanglement between the different bosonic species is also worth studying. This interspecies entanglement in mixtures can provide a new perspective in understanding quantum phase transitions \cite{Wang_2016}. Moreover, a multi-component bosonic mixture would exhibit interesting features in terms of interspecies entanglement for the SCF state. The interspecies entanglement property can be clarified by using the quantum information tool, i.e. the entanglement entropy (EE) \cite{EE1,EE2,EE3,SD}. To calculate the interspecies EE, we use the species partition rather than the space-like partitions (lattice partitions). In addition, we note that since we applying the single-site Gutzwiller method, the interspecies EE in this present study reflects the entanglement properties of the single-site ground state, which is connected to the true ground state. We expect that different phases should have different values of interspecies EE. For instance, the interspecies EE is expected to be zero in the Mott insulator state, as this state is a non-entangled local phase. Conversely, the EE value must be non-zero in a superfluid phase. Unfortunately, although the interspecies EE has been discussed in a two-component bosonic mixture \cite{Wang_2016, SD,EE2CBM2}, there is still no research to study the interspecies entanglement properties for a mixture with more than two components. 

Besides, the different species can not occupy the same site when the interspecies interaction strength is repulsive and sufficiently large which is known as the demixing effect or phase separation \cite{demixing1,demixing2,demixing3,demixing4}. For an interexchange symmetric two-component mixture, people have known that the mixing-demixing phase transition occurs when the interspecies interaction strength is greater than the intraspecies interaction strength \cite{demixing5,demixing6}. But for a mixture with three or more than three components, the investigation of the demixing effect is lacking. In the present work, we study the demixing effect for a $n$-component bosonic mixture by tuning the interspecies interaction strength and show that the critical point of the mixing-demixing phase transition is independent of $n$. 

This paper is organized as follows. First, in \ref{Model}, we describe the model, the numerical method, and the measurements. Then, in \ref{Results}, we present the numerical calculations for a two- and three-component bosonic mixture and generalize these results to a $n$-component mixture. The main ones are: (1) the phase diagram and the order parameters, (2) the interspecies entanglement entropy, (3) the spectral decomposition of the ground state for different phases, and (4) the influence of the demixing effect
on the above calculations. Finally, in \ref{Conclusions}, we provide a brief summary and conclusions.
\section{Model and methods} \label{Model}
We start with the Hamiltonian of a $n$-component ($n=2,3$) Bose-Hubbard model,
\begin{align}\label{Ham}
H=&-J\sum_{\alpha,\langle i,j\rangle}\left(\alpha_i^\dag\alpha_j + h.c.\right)+\sum_{\alpha,i}\frac{U_{\alpha}}{2}n_{i,\alpha}(n_{i,\alpha}-1) \\ \nonumber
&+\sum_{\alpha < \alpha^\prime,i}U_{\alpha,\alpha^\prime}n_{i,\alpha}n_{i,\alpha^\prime}-\mu\sum_{\alpha, i}n_{i,\alpha},
\end{align}
where $\alpha$ labels the $\alpha$th-component bosons,  e. g. $\alpha=a, b$ and $a,b,c$ for a two- and three-component bosonic mixture respectively, $\alpha_i$ is the $\alpha$-boson annihilation operator on site $i$,  $n_{\alpha}$ is the $\alpha$-boson number operator,  and $\langle i,j\rangle$ represents a nearest-neighbor summation. In the Eq. (\ref{Ham}), $J>0$ is the nearest-neighbor hopping amplitude, $U_{\alpha}>0$ is the $\alpha$th-component onsite intraspecies repulsive interaction, $U_{\alpha,\alpha^\prime}>0$ is the $\alpha$-$\alpha^\prime$ onsite interspecies repulsive interaction, and $\mu>0$ is the chemical potential. For a homogeneous system, an interexchange symmetry would be preserved between every two different species, which means $U_\alpha$ is the same for each component and $U_{\alpha,\alpha^\prime}$ is also the same for every two species. For instance, we can set $U_a=U_b=U$ for a two-component bosonic mixture, and $U_a=U_b=U_c=U$, $U_{ab}=U_{bc}=U_{ac}=U^\prime$ for a three-component one. 

To solve Eq. (\ref{Ham}), we use the single-site Gutzwiller approach (SSGA)\cite{SingleGutz1,SingleGutz2,SingleGutz3,SingleGutz4,SingleGutz5,SingleGutz6,SingleGutz7}, which is a mean-field method that assumes the ground state wave function can be written as a product of a single-site Gutzwiller trial wave function $|i\rangle$ and the wave function of all remaining sites $|\psi\rangle$,
\begin{align}\label{WF}
|\Psi\rangle=|i\rangle|\psi\rangle,
\end{align}
where $|i\rangle$ can be represented in a local single-site Fock space, e.g. $|i\rangle$ is equal to $\sum_{m_a,m_b} c_{m_a,m_b}|m_a,m_b\rangle$ and $\sum_{m_a,m_b,m_c} c_{m_a,m_b,m_c}|m_a,m_b,m_c\rangle$ for a two- and three-component bosonic mixture respectively. In general, we can write $|i\rangle$ as,
\begin{align}\label{SinglesiteWF}
|i\rangle=\sum_{\{m_\alpha\}} c_{\{m_\alpha\}}|\{m_\alpha\}\rangle.
\end{align}
Here, $c_{\{m_\alpha\}}$ is the coefficient and $m_\alpha$ ($\alpha=a,b$ or $a,b,c$) is the particle occupation number of $\alpha$-th component. In the spirit of the SSGA, we do not need to know the knowledge of $|\psi\rangle$, and what we do in the method is just use a self-consistent loop diagonalization scheme to obtain the single-site wave function $|i\rangle$ (See details in Appendix A). In other words, we can project the Hamiltonian Eq.(\ref{Ham}) into a local small single-site Fock space and the model is then treated as a single lattice site coupled only to the average mean field. That is, the Hamiltonian in the SSGA can be read as,
\begin{align}\label{HamSSGA}
H_{\mathrm{SSGA}}=&-zJ\sum_{\alpha}\left(\alpha_i^\dag \langle \alpha \rangle + h.c.\right)+\sum_{\alpha}\frac{U_{\alpha}}{2}n_{\alpha}(n_{\alpha}-1) \\ \nonumber
&+\sum_{\alpha < \alpha^\prime}U_{\alpha,\alpha^\prime}n_{\alpha}n_{\alpha^\prime}-\mu\sum_{\alpha}n_{\alpha},
\end{align}
where $z$ is the number of the nearest-neighbor sites of $i$ site. Here, we assume that the bosons are loaded in a two-dimensional square optical lattice, and it means that $z=2d=2\times2=4$ is fixed.  $\langle \alpha \rangle$ is the mean-field parameter which is equal to $\langle i| \alpha |i\rangle$. From Eq.(\ref{HamSSGA}),  we know that although the SSGA is a mean-field method, there are more internal degrees of freedom if we apply it to a multi-component bosonic mixture. For example, if we set the maximum boson occupation number of each component to be $N_\alpha=10$ ($\alpha=a,b,c$) on account of the repulsive intraspecies interaction, then the dimension $D$ of the Hilbert space is $D=(N_a+1) \times (N_b+1)=11\times11=121$ and $D=(N_a+1) \times (N_b+1)\times (N_c+1)=11\times11\times11=1331$ for a two- and three-component mixture respectively. Obviously, $D$ grows exponentially with the number of boson components.  Note that this boson number cutoff is needed in numerical calculations.

The other important parameter in the SSGA is self-consistent convergence accuracy. In our calculations, it is set to be $\Delta<10^{-15}$, where $\Delta$ is defined as,
\begin{align}\label{Delta}
\Delta=|E_{\ell}/U-E_{\ell-1}/U|+\sum_{\alpha}|\langle \alpha\rangle_{\ell} -\langle \alpha \rangle_{\ell-1}|<10^{-15}
\end{align}
which is the energy and SF order parameters difference between two continuous self-consistent iterations. Here, $\ell$ indicates the iteration index within a self-consistent loop. When a self-consistent loop is finished, the single-site wave function $|i\rangle$ and the SF mean-field order parameters $\langle \alpha \rangle$  would be both determined. Using the value of $\langle \alpha \rangle$, the SF-MI phase boundary can be located. Due to this, a straightforward method to obtain the mean-field phase diagram is to calculate $\langle \alpha \rangle$ in the whole $zJ/U$-$\mu/U$ parameter space (here, we set $U$ as the energy unit while varying $J$ and $\mu$).  But we can do this more sufficiently and precisely by involving a binary search algorithm: at each value of $\mu$, the MI-SF critical point  $J^c$ can be found in following ways, i) initializing three $J$ points, i.e. $J_{\mathrm{min}}=0$, $J_{\mathrm{max}}=0.4$, and $J_{\mathrm{mean}}=(J_{\mathrm{min}}+J_{\mathrm{max}})/2$; ii) evaluate the value of $\langle \alpha \rangle$ at $J_{\mathrm{mean}}$; iii) if it is smaller than a value, say $10^{-5}$, then ($J_{\mathrm{mean}}$,$\mu$) is assumed in a MI phase region and  $J_{\mathrm{min}}$ is replaced by $J_{\mathrm{mean}}$,  otherwise $J_{\mathrm{max}}=J_{\mathrm{mean}}$; iv) repeat steps i)-iii), until $J_{\mathrm{max}}-J_{\mathrm{min}}$ smaller than another value, e.g. $10^{-6}$. After this procedure is accomplished, the value of $J^c$ is then approximately equal to $(J_{\mathrm{min}}+J_{\mathrm{max}})/2$.  Applying the binary search algorithm with only about $10\sim 20$ iterations for a given value of $\mu$ allows us to determine the critical value with a relative precision of $10^{-5} \sim 10^{-6}$ \cite{SingleGutz7}. The SCF-SF phase boundary is determined in the same way, but in the SCF phase, $\langle \alpha \rangle=0$ while $\langle \alpha {\alpha^\prime}^\dag \rangle_{\alpha \neq \alpha^\prime}\neq 0 $ ($\alpha=a,b$ and $a,b,c$ for a two- and three-component mixture respectively). 

Apart from the mean-field order parameters, the spectral decomposition of the ground state \cite{SD} also reveals the features of the phases. Based on Eq.(\ref{SinglesiteWF}), the ground state of the SSGA Hamiltonian in Eq.(\ref{HamSSGA}) can be written as,
\begin{align}\label{SinglesiteGS}
|i\rangle_0=\sum_{\{m_\alpha\}} c_{\{m_\alpha\}}^0|\{m_\alpha\}\rangle,
\end{align}
which is an expansion of the single-site Fock states. The spectral decomposition relates to the properties of the single-site state. As one can see, $c_{\{m_\alpha\}}^0$ is a tensor, i.e. $c_{\{m_\alpha\}}^0=c_{m_a,m_b}^0$ and $c_{\{m_\alpha\}}^0=c_{m_a,m_b,m_c}^0$ for a two- and three-component mixture respectively. We can define the spectral decomposition as $|c_{\{m_\alpha\}}^0|^2$, and plot its value in the $\{m_\alpha\}$ space.  Different phases would have different characteristics in the spectral decomposition. For example, the ground state is just a single Fock state $|\{m_{\alpha}\}\rangle$ in a MI phase and only a single sharp peak would be observed.

If we are interested in the entanglement between different species, we can calculate the von Neumann entropy. This interspecies entanglement entropy is evaluated by using the partitions of the species degree of freedom \cite{Wang_2016}. Let us talk about some details here. Suppose the multi-component bosonic mixture under consideration can be divided into two parts, P and Q, where P and Q correspond to different species. Then, the single-site ground state Eq.~(\ref{SinglesiteGS}) can be rewritten as $\sum_{m_P,m_Q} C_{m_P,m_Q}^0|m_P,m_Q\rangle$, where $m_P$ and $m_Q$ are the particle numbers of their respective parts. Here, $C_{m_P,m_Q}^0$ denotes the new reshaped matrix elements. We then perform Singular Value Decomposition (SVD) on $C^0$, yielding $C^0=UDV^\dag$. The square diagonal matrix $D$ contains non-negative singular values $\lambda_k$, with its dimension being $d=\min\{\mathrm{Dim}(P),\mathrm{Dim}(Q)\}$. Using the SVD of $C^0$, we can get the Schmidt decomposition of $|i\rangle_0$,
\begin{align}\label{SinglesiteGS1}
|i\rangle_0=\sum_{k=1}^{d} \lambda_k |\psi_P\rangle_k |\psi_Q\rangle_k.
\end{align}
The density matrix (DM) is defined as $\rho=|i\rangle_0 {}_0\langle i |$, and thus the reduced DM of $P$ part is obtained by tracing out the Q part: $\rho_P=\mathrm{Tr}_Q\rho$. Using the Eq.~(\ref{SinglesiteGS1}), we know that the $\rho_P$ is already diagonal with the help of SVD and we can directly write the interspecies EE in terms of $\lambda_k$, 
\begin{align}\label{EE}
S^P=-\operatorname{Tr}_P\left[\rho_P \ln \left(\rho_P\right)\right]=-\sum_k |\lambda_k|^2 \ln |\lambda_k|^2,
\end{align}
where $\ln(\cdots)$ is the natural logarithm. We remind that we perform the SVD on the new reshaped matrix $C_{\{m_P,m_Q\}}^0$ not on the original coefficient tensor $c^0_{\{m_a,m_b,\cdots\}}$. $C_{\{m_P,m_Q\}}^0$ is our start point for calculating the EE.

Because $\lambda_k$ comes from the SVD of the single-site ground state $|i\rangle_0$, Eq.~(\ref{EE}) therefore reveals the interspecies entanglement properties of the single-site ground state which is connected to the entire true system. If $\lambda_k=0$, then the limit of the corresponding term in Eq.~(\ref{EE}) is zero and it is not included in the summation. Besides, the minimum value of $S^P$ is zero when all of the $\lambda_k$ equals zero except $\lambda_{k_0}=1$. On the other side,  the maximum value of $S^P$ can be reached under the condition that
\begin{align}\label{maxEE}
|\lambda_1|^2=|\lambda_2|^2=\cdots = |\lambda_d|^2, 
\end{align}
as reported in \cite{SD}. Furthermore, from Eq.~(\ref{SinglesiteGS1}) and Eq.~(\ref{EE}), we can know that $S^P=S^Q$ due to the singular values are the same whenever traced out $P$ or $Q$ part.  For example, in a three-component mixture, when we chose the partition $P=a$ and $Q=b c$, then $S^a$ is equal to $S^{bc}$. Since the interexchange symmetry is preserved, all the three component bosons are equivalent, and that means $S^{a}=S^{b}=S^{c}$. As a result,
\begin{align}\label{EEthree}
S^a=S^b=S^c=S^{ab}=S^{bc}=S^{ac}.
\end{align}
However, if the interexchange symmetry is broken, then $S^{a}\neq S^{b} \neq S^{c}$ but $S^{a}$ is still equal to $S^{bc}$ ($S^{b}=S^{ac}$ and $S^{c}=S^{ab}$ also hold).

For a four-component bosonic mixture, the partition choice can be $P=a$ ($Q=bcd$) and $P=ab$ ($Q=cd$). If we chose $P=a$ and $Q=bcd$, then $S^a=S^{bcd}=S_1$. While if we chose $P=ab$ and $Q=cd$, then $S^{ab}=S^{cd}=S_2$. In this case, $S_1 \neq S_2$. For simplicity, we only consider the entanglement between a single species and all the other species in this paper, which implies that the partition choice is $P=a$ and $Q=bcd\cdots$. 

\section{Numerical results} \label{Results} 
In the following, we give the results of the measurements which we have discussed above. To study the demixing effect, the evolution of the observables with a tunable value of $U_{ab}$ (in units of $U$) is calculated.  We show the numerical results for a two- and three-component mixture in subsection \ref{2CBM} and \ref{3CBM} respectively, then we generalize it to a $n$-component mixture in subsection \ref{nCBM}. We recall that all the ground-state properties presented below are calculated based on the obtained single-site ground state.

\subsection{Ground state properties of a two-component bosonic mixture} \label{2CBM}
\begin{figure*}[htbp] 
\centering
\includegraphics[width=0.9\textwidth]{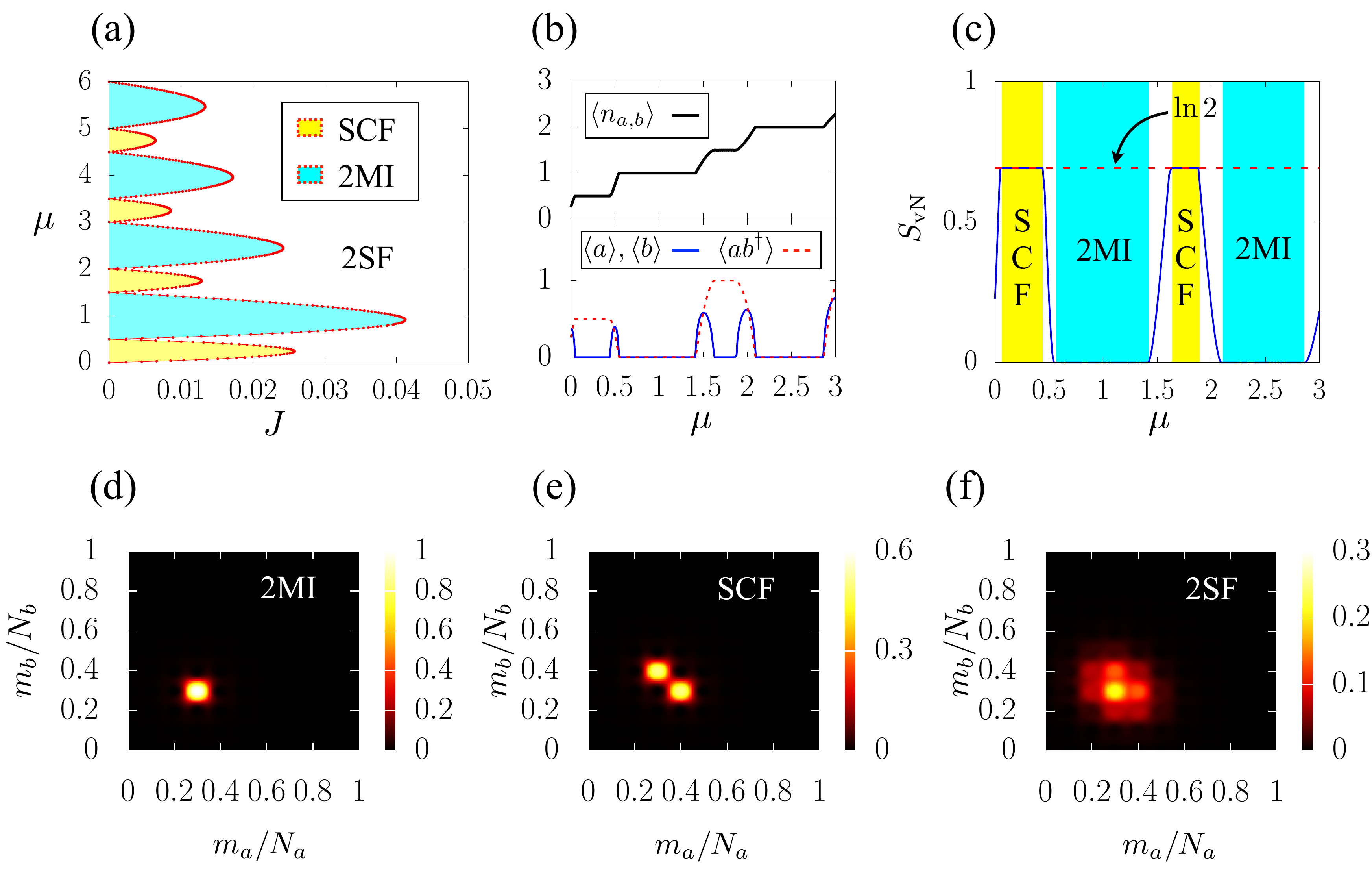}
\caption{ \label{Fig1}(Color online) Ground state properties of a two-component bosonic mixture. The model parameters are set as $U_{a}=U_{b}=U$ and $U_{ab}/U=0.5$. (a) The $zJ/U-\mu/U$ phase diagram (in units of $U$). We use an index $m=1,2,\cdots$ to label the first, second, ... 2MI and SCF lobes. Once $m$ is determined, the ground state of the corresponding lobe is addressed. (b)-(c) The order parameters and the single-site interspecies EE as a function of $\mu/U$ for a fixed value $zJ/U=0.04$. In (c), the white regions (uncolored) indicate the 2SF phase. (d)-(f) $|c_{m_a,m_b}|^2$ for three distinct phases.  $N_a=N_b=10$ is the maximum boson occupation number of component $a$ and $b$, respectively. (d) The spectral decompositions of the third 2MI lobe ($m=3$) for $(zJ/U,\mu/U)=(0.04, 4.0)$. (e) The spectral decompositions of the fourth SCF lobe ($m=4$) for $(zJ/U,\mu/U)=(0.016, 4.8)$. (f) The spectral decompositions of the 2SF phase for $(zJ/U,\mu/U)=(0.4, 4.0)$. In (d), a single peak is observed at $(0.3,0.3)$, representing the 2MI ground state is $|3,3\rangle$. In (e), the two peaks are present at $(0.3,0.4)$ and $(0.4,0.3)$, and its ground state is thus $(|4,3\rangle+|3,4\rangle)/\sqrt{2}$. In (f), there are several peaks that reveal a 2SF phase. Also, we use an interpolation algorithm to make the data more smooth in Figs. (d)-(f).  }
\end{figure*}
The phases of a two-component bosonic mixture have been widely investigated\cite{SingleGutz6, SD, 2CBM1, 2CBM2,2CBMPhase1,2CBMPhase2,2CBMPhase3, 2CBM6,2CBM7,2CBM8,2CBM9}. To conveniently study the properties of different phases, we give the ground state phase diagram under the parameters $U_a=U_b=U$ and $U_{ab}/U=0.5$ in Fig.~\ref{Fig1} (a). It can be seen that there are three phases in the $zJ/U-\mu/U$ phase diagram, namely 2MI, SCF, and 2SF. The corresponding order parameters are shown in Fig.~\ref{Fig1} (b). Fig.~\ref{Fig1} (a) shows the first four 2MI and SCF lobes. We can use an index $m$ to label each 2MI lobe and SCF lobe. For example, $m=1,2,\cdots$ refers to the first, second, ... 2MI lobes and SCF lobes. Once $m$ is determined, the ground state of the corresponding lobe can be known. For instance, the ground state of the $m$-th 2MI and SCF lobe is $|\psi_{\mathrm{2MI}}\rangle=|m_a,m_b\rangle = |m,m\rangle$ and $|\psi_{\mathrm{SCF}}\rangle= (|m,m-1\rangle + |m-1,m\rangle)/\sqrt{2}$, respectively. 

In the SSGA, the 2MI and 2SF are both non-degenerate, but the ground state of the SCF phase is doubly degenerate, e.g. $|1,0\rangle$ and $|0,1\rangle$ are the two degenerate ground states in the first SCF lobe of Fig.~\ref{Fig1} (a). The true ground state of the first SCF lobe is a symmetric summation of the two degenerate states: $|\psi\rangle_0=(|1,0\rangle+|0,1\rangle)/\sqrt{2}$. This degeneracy arises from the fact that a boson from one component should be paired with the boson hole from another component in an SCF state, and it leads to the two possible states ($|1,0\rangle$ and $|0,1\rangle$) satisfying this particle-hole pairing. Thus, in the first SCF lobe,  $\langle a \rangle=\langle b \rangle=0$, $\langle ab^\dag \rangle = 0.5$, and $\langle n_a \rangle=\langle n_b\rangle=0.5$ as shown in Fig.~\ref{Fig1} (b). All these results are in good agreement with the previous studies. 

One of our main goals is to calculate the interspecies EE, which is an important feature of a bosonic mixture. Let us consider it for the 2MI phase first. From its single-site ground state, we can know that the interspecies EE is zero in all the 2MI lobes, indicating that the 2MI phase is a non-entangled state. While for the SCF phase, the interspecies EE is nonzero. The two singular values of the SVD in the $m$-th SCF lobe are $\lambda_1=\lambda_2=1/\sqrt{2}$. It is immediately known that the interspecies EE in the SCF phase is a constant $\ln2$, representing the maximum value reached. For the 2SF phase, the ground state is a linear combination of the Fock states, and the value of the coefficients $c_{\{m_\alpha\}}$ depends on the parameters, meaning that the interspecies EE is evolved with $\mu$.  All these predictions are observed in Fig.~\ref{Fig1}(c).

In addition, spectral decomposition is the other feature of the ground state. In Figs.~\ref{Fig1} (d)-(f), we present the spectral decomposition for the third 2MI lobe ($m=3$), the fourth SCF lobe ($m=4$), and the 2SF phase in the same order as listed. Specifically, the ground state of the third 2MI lobe ($m=3$) and the fourth SCF lobe ($m=4$) is $|3,3\rangle$ and $(|4,3\rangle+|3,4\rangle)/\sqrt{2}$ respectively, resulting in a single sharp peak in Fig.~\ref{Fig1} (d) and two peaks in Fig.~\ref{Fig1} (e). But for the 2SF phase, which has a ground state that is a linear combination of some Fock states, several broadening peaks are found in  Fig.~\ref{Fig1} (f).

\begin{figure}[htbp]
\centering
\includegraphics[width=0.45\textwidth]{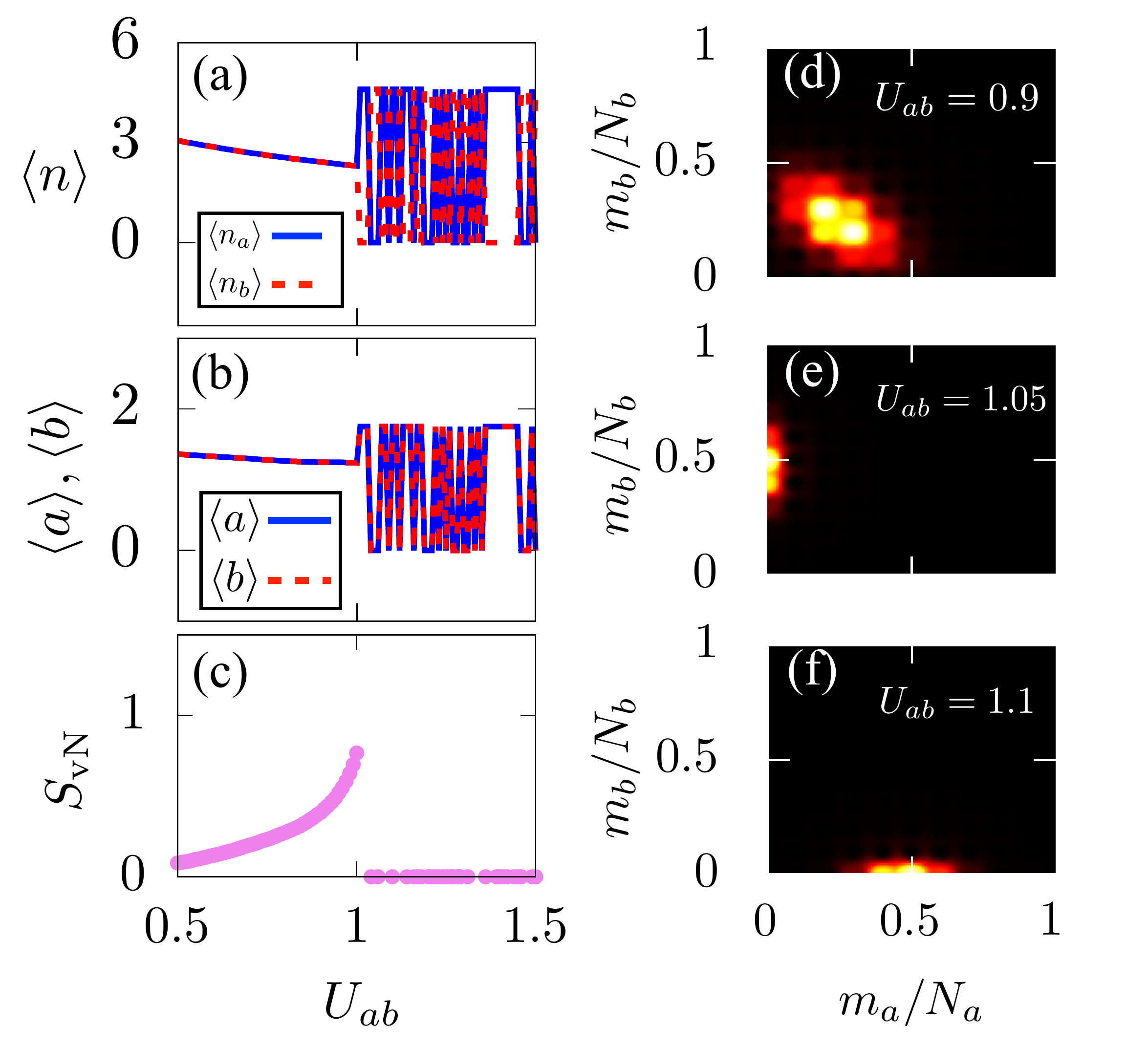}
\caption{\label{Fig2}(Color online) Demixing effect of the 2SF phase for a two-component bosonic mixture under the parameters $zJ/U=0.4$ and $\mu/U=4.0$.  (a)-(b) The averaged particle numbers and the order parameters as a function of $U_{ab}/U$. (c) The interspecies EE as a function of $U_{ab}/U$. (d)-(f) The spectral decompositions for three different values of $U_{ab}/U$.  Here, we set $U$ as the energy unit. $N_a=N_b=10$ is the maximum boson occupation number of component $a$ and $b$, respectively.}
\end{figure}

Now, let us examine what happened at a sufficiently large interspecies interaction strength. People have studied this situation and found that a demixing effect (phase separation) occurs when $U_{ab}/U>1$ \cite{demixing5,demixing6,demixing7,demixing8}. This critical point depends on the particle filling factors \cite{demixing7,demixing8}. Using the SSGA, we can also reproduce this demixing effect.  In Figs.~\ref{Fig2}(a)-(b), we present the mean-field SF order parameters and the averaged particle numbers as a function of $U_{ab}$ (in units of $U$). The figures clearly show that the initial phase at $U_{ab}/U=0.5$ is a mixed 2SF state because the value of $\langle a \rangle$ and $\langle b \rangle$ are both non-zero. Subsequently, an unstable phase has taken place when $U_{ab}/U>1$ since the locations of sudden changes in Fig.~\ref{Fig2}(a)-(b) appear random. This random behavior results from the fact that the bosons of different species can never coexist at the same site in the demixing state, and it leads to the single site being randomly occupied by $a$ and $b$ bosons. It is a limitation of the method used, as the SSGA only considers bosons at a single site. For that reason, the locations of sudden changes hold no physical meanings. However, the nonzero values of the corresponding measurements are not random and independent of the initial conditions of the numerical method. For example, in Fig.~\ref{Fig2}(a), $\langle n_{a} \rangle$ ($\langle n_{b} \rangle$) equals to a constant 4.597 at $U_{ab}/U>1$ when $\langle n_{b} \rangle$ ($\langle n_{a} \rangle$) is zero (See details in Appendix B). This special value of 4.597 depends on the Hamiltonian parameters rather than the self-consistent process, which makes it meaningful.

The mixed-demixed phase transition can also be seen in the evolution of the interspecies EE. As shown in Fig.~\ref{Fig2}(c), $S$ increases as $U_{ab}/U$ goes from 0.5 to 1.0 and reaches a large value at $U_{ab}/U=1.0$. This behavior owing to the increasing interspecies pairing strength as the value of $U_{ab}/U$ increases. But if $U_{ab}/U$ is too strong, e.g. $U_{ab}/U>1.0$, then the mixture is in a demixed phase and $S$ is zero because the lattice site only can be occupied by the single species. The zero value also reflects that each component is not entangled with any other component in such a demixed phase.

To illustrate how the demixing effect affects the spectral decompositions, we show it for three different values of $U_{ab}/U$ in Figs.~\ref{Fig2}(d)-(f). The features in Fig.~\ref{Fig2}(d) are similar to that in Fig.~\ref{Fig1}(f), which indicates a mixed 2SF phase. Once the demixing effect takes over, only a single component of bosons can be observed in the ground state. That is what we have shown in the figures. For example, only the $b$ and $a$ component bosons are observed in Fig.~\ref{Fig2}(e) and (f) respectively. 

\subsection{Ground state properties of a three-component bosonic mixture} \label{3CBM}
\begin{figure*}[htbp] 
\centering
\includegraphics[width=0.9\textwidth]{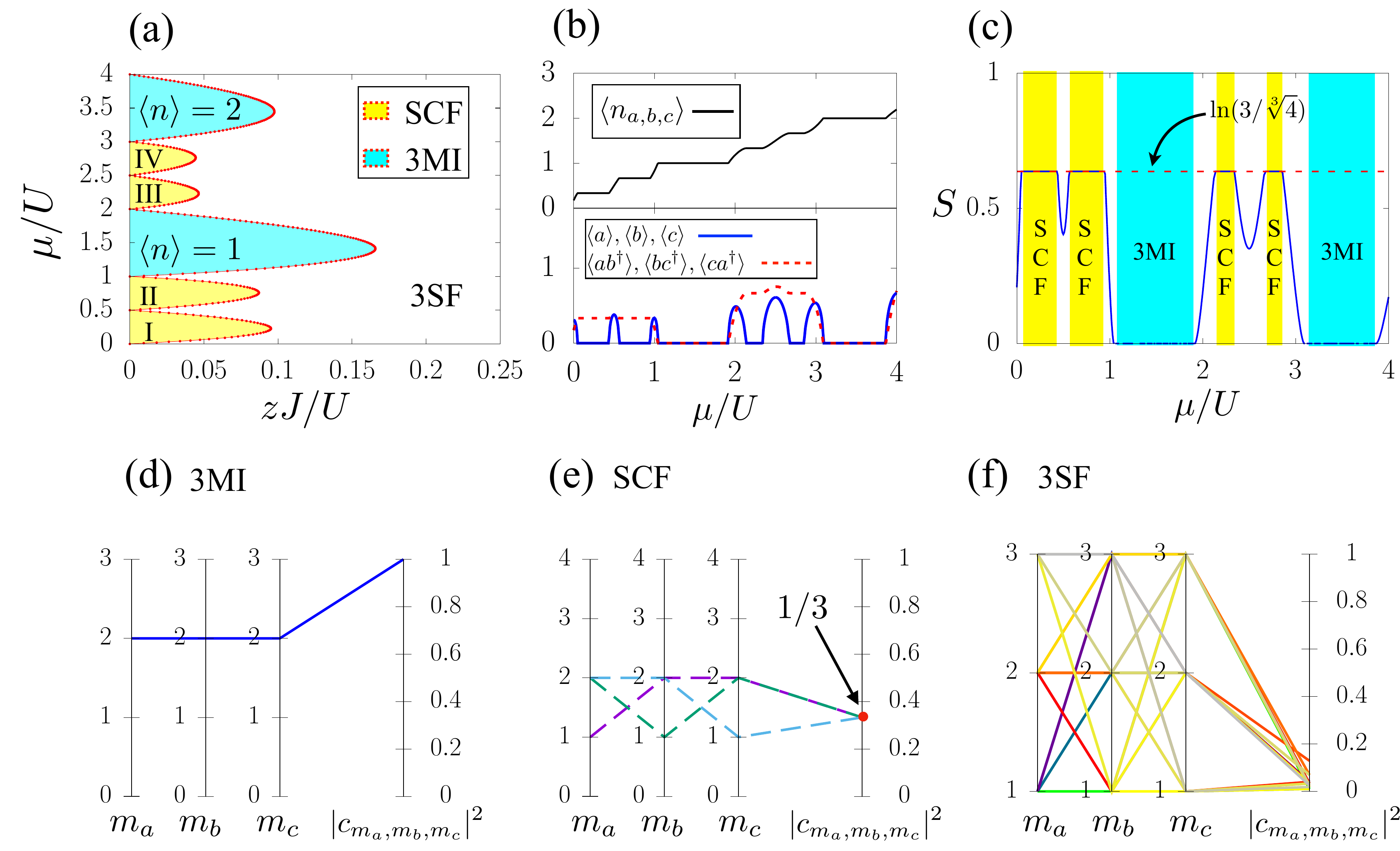}
\caption{ \label{Fig3}(Color online) Ground state properties of a three-component bosonic mixture. The model parameters are set as $U_{a}=U_{b}=U_c=U$ and $U_{ab}/U=U_{bc}/U=U_{ac}/U=U^\prime/U=0.5$. (a) The $zJ/U-\mu/U$ phase diagram (in units of $U$). We use an index $m$ to label the first, second, ... 3MI lobes and use $(m,k)$ to label the SCF lobes ($m=1,2,3,\cdots$ and $k=1,2$). In the SCF-I, II, III, and IV lobes, $(m,k)$ is $(1,1)$, $(1,2)$, $(2,1)$, and $(2,2)$ respectively. Once the indices are determined, the ground state of the corresponding lobe is addressed. (b)-(c) The order parameters and the interspecies EE as a function of $\mu/U$ for a fixed value $zJ/U=0.04$.  In (c), the white regions (uncolored) indicate the 3SF phase. (d)-(f) $|c_{m_a,m_b,m_c}|^2$ for three different phases.  In (d)-(f), the parallel coordinates plot is used to display the spectral decompositions, which means the variables $m_a$, $m_b$, $m_c$, and $|c_{m_a,m_b,m_c}|^2$ are represented by a vertical axis, respectively. A line is drawn connecting the values of each variable for each data point. (d) The spectral decompositions of the second 3MI lobe ( $m=2$ ) for $(zJ/U, \mu/U) = (0.04, 3.5)$. The ground state of this lobe is known as $|2,2,2\rangle$, and it is characterized by a single peak in the figure. (e) The spectral decompositions of the SCF-IV ($(m,k)$=$(2,2)$) lobe for $(zJ/U, \mu/U) = (0.02, 2.7)$. Its ground state is $(|1,2,2\rangle+|2,1,2\rangle+|2,2,1\rangle)/\sqrt{3}$, symbolized by three dashed lines in the figure. (f) The spectral decompositions of the 3SF phase for $(zJ/U, \mu/U)=(0.4, 3.5)$. The ground state in this phase is a linear combination of several Fock states, which is why multiple lines can be observed. }
\end{figure*}
We recall that we consider a homogenous three-component bosonic mixture here for simplicity. This means that we can set the model parameters to be  $U_a=U_b=U_c=U$ and $U_{ab}/U=U_{bc}/U=U_{ac}/U=U^\prime/U=0.5$. The $zJ/U-\mu/U$ phase diagram for this three-component bosonic mixture is given in Fig.~\ref{Fig3}(a).  Comparing the phase diagram of the two-component bosonic mixture in Fig.~\ref{Fig2}(a), we find, to our surprise, that two SCF lobes instead of a single SCF lobe separate the neighboring MI lobes. The reason is that the ground state of the SCF phase for a three-component bosonic mixture has two different forms. To label these SCF lobes, a single index $m$ is insufficient, and an additional index $k$ is required. We thus use the notation $(m,k)$ to represent the SCF lobes of Fig.~\ref{Fig3}(a), where $m=1,2,3\cdots$ and $k=1,2$. For convenience, we give the general expression of these two kinds of SCF states here.  If $k=1$, the SCF ground state is,
\begin{align}\label{SCF1}
|\psi\rangle_{\mathrm{SCF-(1)}}^m&=\frac{1}{\sqrt{3}}(|m,m-1,m-1\rangle \nonumber\\ 
&+|m-1,m,m-1\rangle+|m-1,m-1,m\rangle ).
\end{align}
And if $k=2$, the SCF ground state is,
\begin{align}\label{SCF2}
|\psi\rangle_{\mathrm{SCF-(2)}}^m&=\frac{1}{\sqrt{3}}(|m,m,m-1\rangle \nonumber\\ 
&+|m,m-1,m\rangle+|m-1,m,m\rangle ).
\end{align}
Once $(m,k)$ is determined, the ground state of the corresponding SCF lobe is identified. 

To illustrate this, we give some examples. In Fig.~\ref{Fig3}(a), we show the first four SCF lobes. For the SCF-I (II) lobe, $m=1$ and $k=1 (2)$, then its ground state is $|\psi\rangle_{\mathrm{SCF-(1)}}^1$ ($|\psi\rangle_{\mathrm{SCF-(2)}}^1$). For the SCF-III (IV) lobe, $m=2$ and $k=1 (2)$, the corresponding ground state is $|\psi\rangle_{\mathrm{SCF-(1)}}^2$ ($|\psi\rangle_{\mathrm{SCF-(2)}}^2$). Moreover, the SCF order parameters $\langle ab^\dag \rangle$, $\langle bc^\dag \rangle$, and $\langle ca^\dag \rangle$ in the $(m,k)$-th SCF lobe are all equal to $m/3$, which is independent of $k$, e.g. in the SCF-I and SCF-II lobes, $\langle ab^\dag \rangle$ is equal to 1/3. All these predictions are examined in Fig.~\ref{Fig3}(b). Eq.~(\ref{SCF1}) and Eq.~(\ref{SCF2}) are similar to that of the two-component mixture case, as they are both symmetric summations of Fock states. This is because these Fock states become degenerate states when the mixture is situated within an SCF lobe. On account of this, the SCF ground state degeneracy is three for a three-component mixture. On the other hand, we can still use a single index $m$ to label the 3MI lobes with the ground state given as,
\begin{align}\label{3MI}
|\psi\rangle_{\mathrm{3MI}}^m&=|m,m,m\rangle, \quad m=1,2,3,\cdots.
\end{align}

Let us now calculate the average particle numbers in the SCF phase. It is easy to obtain that $\langle n_{\alpha} \rangle$ ($\alpha=a,b,c$)  are $m-2/3$ and $m-1/3$ in Eq.~(\ref{SCF1}) and Eq.~(\ref{SCF2}), respectively. For instance, if we set $m=1$, then $\langle n_{\alpha} \rangle$ equals 1/3 in Eq.~(\ref{SCF1}) and 2/3 in Eq.~(\ref{SCF2}), corresponding to the SCF-I and SCF-II phases, respectively. Similarly, if $m=2$, then $\langle n_{\alpha} \rangle=4/3$ in Eq.~(\ref{SCF1}) and 5/3 in Eq.~(\ref{SCF2}), associated with the SCF-III and SCF-IV lobe, respectively. This has been examined by the numerical calculations in Fig.~\ref{Fig3}(b). Due to these special average particle numbers,  every two neighboring 3MI lobes with $\langle n_{\alpha} \rangle$ being an integer $m-1$ and $m$ should be separated by two SCF lobes. For example, the first 3MI lobe with its ground state being $|1,1,1\rangle$ is separated from the second 3MI lobe with its ground state being $|2,2,2\rangle$ by SCF-III and SCF-IV lobes. This has been shown in Fig.~\ref{Fig3}(a).

The interspecies EE of a three-component bosonic mixture is also somewhat different from that of a two-component mixture. To calculate its values, we can still divide the three-component mixture into a bipartite system, i.e. by tracing out the degrees of freedom of $bc$ species and getting the interspecies EE between $a$ and $bc$ bosons. The result is presented in Fig.~\ref{Fig3}(c). As it is shown,  $S=0$ in a 3MI state because it is a non-entangled phase, which is the same as that in a 2MI state. But for an SCF state, $S$ is equal to $\ln (3 / \sqrt[3]{4})$ rather than $\ln 2$. We can understand it by performing an analytical calculation of $S$. We treat the SCF ground state Eq.~(\ref{SCF1}) as a tensor $c$, 
\begin{align}\label{Ctensor}
c_{m-1,m-1,m}=c_{m-1,m,m-1} =c_{m,m-1,m-1}=\frac{1}{\sqrt{3}},
\end{align}
where all other elements are zero.  The latter two indices correspond to the $bc$ components which can be contracted as a single index. For this reason, tensor $c$ can be reshaped to a matrix,
\begin{align}\label{Cmatrix}
c_{m-1,3}=c_{m-1,2} =c_{m,1}=\frac{1}{\sqrt{3}},
\end{align}
and all of the other matrix elements are zero. The reduced DM after tracing out the degrees of freedom of $bc$ bosons is given by
\begin{align}\label{rhoa}
\rho_a=cc^\dag=\left( \begin{matrix} 2/3 & 0 \\ 0 & 1/3 \end{matrix}\right).
\end{align}
The interspecies EE is then given as 
\begin{align}\label{Sthree}
S=-\frac{2}{3}\ln (2/3)-\frac{1}{3}\ln (1/3) = \ln (3 / \sqrt[3]{4}),
\end{align}
If we use the second type of the SCF ground state Eq.~(\ref{SCF2}), we get the same result because the tensor $c$ is the same. Due to this reason, the value of $S$ is the same for all the SCF lobes. Note that $\ln (3 / \sqrt[3]{4})$ is not the maximum value of $S$ here. 

To show the differences between the three phases, we also present the spectral decomposition of the ground states in Figs.~\ref{Fig3}(d)-(f) using the parallel coordinates plot. In Figs.~\ref{Fig3}(d)-(f), the variables $m_a$, $m_b$, $m_c$, and $|c_{m_a,m_b,m_c}|^2$ are represented by a vertical axis, respectively. In the parallel coordinates plots, a line is drawn connecting the values of each variable for each data point.  For the $m$-th 3MI lobe, the ground state is $|m,m,m\rangle$, and thus a single line is seen in Fig.~\ref{Fig3}(d) with $|c_{m,m,m}|^2=1$.  For the $(m,k)$-th SCF lobe, the ground state consists of three Fock states. For example, when $m=2$ and $k=2$, the corresponding ground state is $(|1,2,2\rangle+|2,1,2\rangle+|2,2,1\rangle)/\sqrt{3}$, represented by three dashed lines in Fig.~\ref{Fig3}(e). In the 3SF phase, the ground state is a linear combination of several Fock states, which is why multiple lines can be observed in Fig.~\ref{Fig3}(f). 

Similar to the two-component mixture case, there should also exist the demixing effect in a three-component mixture when $U^\prime$ is sufficiently large. Before discussing the numerical results, we clarify the conditions for the demixing effect in a three-component mixture with the strong coupling limit ($\frac{U^\prime}{J}, \frac{U}{J}\rightarrow \infty$). Here, we consider a zero-hopping $J\rightarrow 0$ situation to get a qualitative understanding of the mixing-demixing critical point.  At $J=0$, the SSGA Hamiltonian Eq.~(\ref{HamSSGA}) for a three-component mixture is read as,
\begin{align}\label{HamSSGA1}
\widetilde{H}_{\mathrm{SSGA}}=&\frac{U}{2}[n_a(n_a-1)+n_b(n_b-1)+n_c(n_c-1)] \nonumber \\
&+U^\prime(n_a n_b+n_b n_c+n_a n_c) \nonumber \\
&-\mu(n_a+n_b+n_c).
\end{align}
As we discussed above, there are two phases at $J=0$, one is the 3MI and the other is SCF. Their ground states are known, and the energies for the two phases can be expressed as,
\begin{align}\label{E3MI}
E_{\mathrm{3MI}}=\frac{U}{2}3m(m-1)+3U^\prime m^2-\mu 3m,
\end{align}
\begin{align}\label{ESCF1}
E_{\mathrm{SCF-(1)}}=&\frac{U}{2}\left[3(m-1)(m-2)+2(m-1)\right] \nonumber\\&+U^\prime [2m(m-1)+(m-1)^2]\nonumber\\&-\mu (3m-2),
\end{align}
\begin{align}\label{ESCF2}
E_{\mathrm{SCF-(2)}}=&\frac{U}{2}\left[3(m-1)(m-2)+4(m-1)\right] \nonumber\\&+U^\prime [2m(m-1)+m^2]\nonumber\\&-\mu (3m-1).
\end{align}

In the limit $U^\prime/U \rightarrow +\infty$, the interspecies interaction becomes infinitely large, which means a single site can not be occupied by different species. This leads to the formation of a demixed state. In the SSGA, the single-site demixed ground state is $|m_a,0,0\rangle$, $|0,m_b,0\rangle$ or $|0,0,m_c\rangle$. Due to the interexchange symmetry being considered, the three demixed states are equivalent. We can analyze the demixing effect by choosing one of the three demixed states arbitrarily, e.g. using $|m_a,0,0\rangle$. Because the total particle number should be conserved during the mixing-demixing phase transition, that is, 3MI phase $\rightarrow |3m,0,0\rangle$, SCF phase $\rightarrow |3m-1,0,0\rangle$ or $|3m-2,0,0\rangle$, after the demixing effect occurs.  For these demixed states to be the ground states, one must satisfy
\begin{align}\label{conditions}
E_{3m,0,0}&<E_{\mathrm{3MI}}, \nonumber \\
E_{3m-2,0,0}&<E_{\mathrm{SCF-(1)}}, \nonumber \\
E_{3m-1,0,0}&<E_{\mathrm{SCF-(2)}},
\end{align}
where $E_{m_a,m_b,m_c}$ is denoted as the energy of a state $|m_a,m_b,m_c\rangle$.  Using these relations and Eqs.~(\ref{E3MI})-(\ref{ESCF2}), we obtain the condition of the demixing effect in the SCF and 3MI phase, $U^\prime>U$, which is the same for a two-component mixture. 

\begin{figure}[htbp]
\centering
\includegraphics[width=0.45\textwidth]{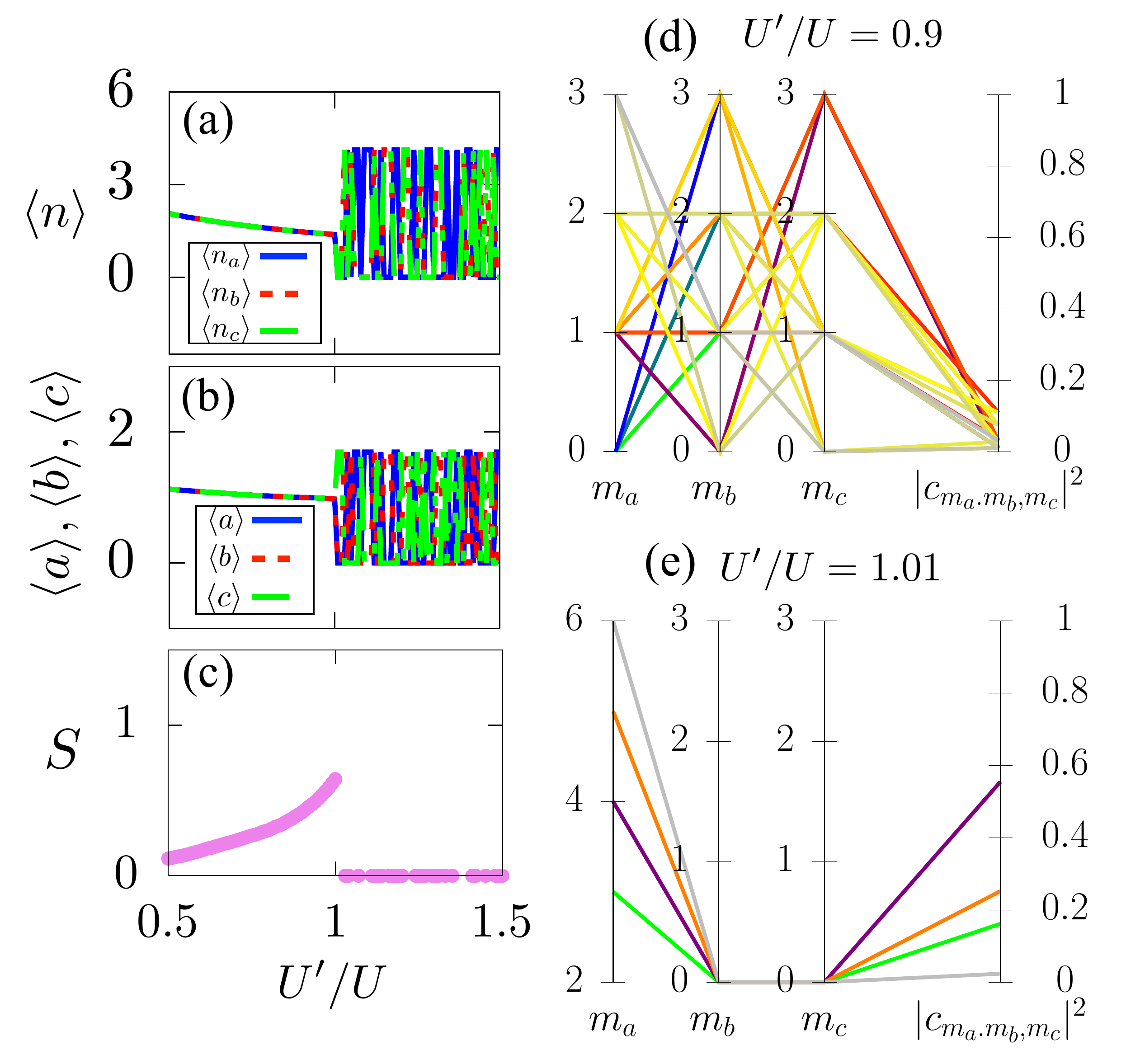}
\caption{\label{Fig4}(Color online) Demixing effect of the 3SF phase for a three-component bosonic mixture under the parameters $zJ/U=0.2$ and $\mu/U=3.5$ (In units of $U$).  (a) and (b) shows the averaged particle numbers and the order parameters as a function of the interspecies interaction strength $U^\prime/U$ respectively. (c) The interspecies EE as a function of $U^\prime/U$. (d)-(e) The spectral decompositions of the ground state wave function for two different values of $U^\prime/U$. In (d), a 3SF phase is formed, and its ground state is a linear combination of several Fock states. In (e), a demixed phase is established and only boson $a$ is observed.}
\end{figure}

For the demixing effect in a 3SF phase, we can analyze it numerically by tuning the value of $U^\prime$ (in units of $U$). The results are presented in Fig.~\ref{Fig4}. As Figs.~\ref{Fig4}(a)-(b) show, the non-zero SF order parameters and fractional averaged particle numbers imply that the mixture is indeed in a 3SF phase when  $U^\prime/U<1.0$. While if $U^\prime/U>1.0$, a demixed phase is established and the measurements in Figs.~\ref{Fig4}(a)-(b) are randomly equal to zero. This suggests that the critical point $U^\prime_c=U$, which is analytically derived for the 3MI and SCF phases at $J=0$, is still valid for the 3SF phase at a finite value of $J$. 

To study the influence of the demixing effect on the interspecies entanglement, $S$ as a function of $U^\prime/U$ is addressed in Fig.~\ref{Fig4}(c). Comparing it to Fig.~\ref{Fig2}(c), we find that the behavior is the same as that in a two-component mixture, e.g. the value of $S$ both increase at $U^\prime/U<1$ and then turn into zero in the demixed state. This is because the increased interspecies repulsive interaction leads to the increased interspecies particle-hole paring strength. Consequently, interspecies entanglement is increased. But if $U^\prime/U$ is too strong, e.g. $U^\prime/U>1$, the mixture would be in a demixed phase. This evolution occurs regardless of whether the system is a two- or three-component mixture. 

Besides, as we do in the two-component mixture, the mixed-demixed phase transition can also be revealed in the spectral decomposition of the ground state. To demonstrate this, we set the parameters $zJ/U=0.2$ and $\mu/U=3.5$. When $U^\prime/U=0.9$, the mixture is in a mixed 3SF phase, and the ground state consists of the Fock states from all three components of bosons. However, if $U^\prime/U>1$, the mixture is in a demixed phase, and only one of the three species is left in its ground state. These differences can be observed in Figs.~\ref{Fig4}(d)-(e).  

\subsection{Generalize to a bosonic mixture with $n$ components} \label{nCBM}
We now generalize the above results to a $n$-component bosonic mixture, where the interexchange symmetry ($a\leftrightarrow b\leftrightarrow c \leftrightarrow d \cdots$) is preserved.  From the discussions in Section \ref{2CBM} and \ref{3CBM}, we can know that there should have three phases in the phase diagram of a $n$-component mixture ($n=2,3,4,\cdots$). For convenience, we label the three phases as nSF, nMI, and SCF.  For simplicity, let us first consider the ground state and energy of the nMI phase.  Based on the numerical calculations of a two- and three-component mixture, the single-site ground state of the $m$-th nMI lobe can be written as
\begin{align}\label{nMI}
|\psi\rangle_{\mathrm{nMI}}^{m}=|m_a,m_b,\cdots,m_n\rangle=|m,m,\cdots,m\rangle,
\end{align}
where the averaged particle number in this state is $\langle n_\alpha \rangle=m$ ($m=1,2,3,\cdots$).  The energy of this nMI state is 
\begin{align}\label{EnMI}
E_{\mathrm{nMI}}^{m}=\frac{U}{2}nm(m-1)+\frac{n(n-1)}{2}U^\prime m^2-\mu nm.
\end{align}

On the other hand, there should exist $n-1$ SCF lobes between the $(m-1)$-th and $m$-th nMI lobe according to the numerical phase diagrams.  We also use the notation $(m,k)$ to label these SCF lobes, where $m=1,2,3\cdots$ and $k=1,2,\cdots,n-1$. According to Section \ref{2CBM} and \ref{3CBM}, the single-site ground state of the $(m,k)$-th SCF lobe can be written as a symmetric summation of the degenerated states
\begin{equation}\label{SCFk}
\begin{aligned}
|\psi\rangle_{\mathrm{SCF-(k)}}^{m}=&\frac{1}{\sqrt{C_n^k}}(| \underbrace{m-1, \cdots, m-1}_{n-k}, \underbrace{m,\cdots,m}_{k}\rangle \\
&+\cdots+ |\underbrace{m,\cdots,m}_k,\underbrace{m-1,\cdots,m-1}_{n-k}\rangle), \\
\end{aligned}
\end{equation}
where $C_n^k=\left(\begin{array}{l}n \\ k\end{array}\right)=\frac{n!}{k!(n-k)!}$ ($k=1,2,3,\cdots, n-1$) is the ground state degeneracy,  and there are $C_n^k$ terms in the summation $(\cdots)$. 

The averaged particle numbers $\langle n_\alpha \rangle$ in the $(m,k)$-th SCF lobe can now be calculated and it is
\begin{equation}\label{<n>}
\begin{aligned}
\langle n_\alpha \rangle_{\mathrm{SCF-(k)}}^m&=m-1+\frac{k}{n},
\end{aligned}
\end{equation}
The energy of this $(m,k)$-th SCF lobe can be also obtained, which is given as
\begin{equation}\label{ESCF}
\begin{aligned}
E_{\mathrm{SCF}-(k)}^m=&\frac{U}{2}[n(m-1)(m-2)+2k(m-1)] \\
&+U^\prime[\frac{(n-k)(n-k-1)}{2}(m-1)^2\\
&+(n-k)km(m-1)+\frac{k(k-1)}{2}m^2]\\
&-\mu[n(m-1)+k].
\end{aligned}
\end{equation}

\begin{figure}[htbp]
\centering
\includegraphics[width=0.45\textwidth]{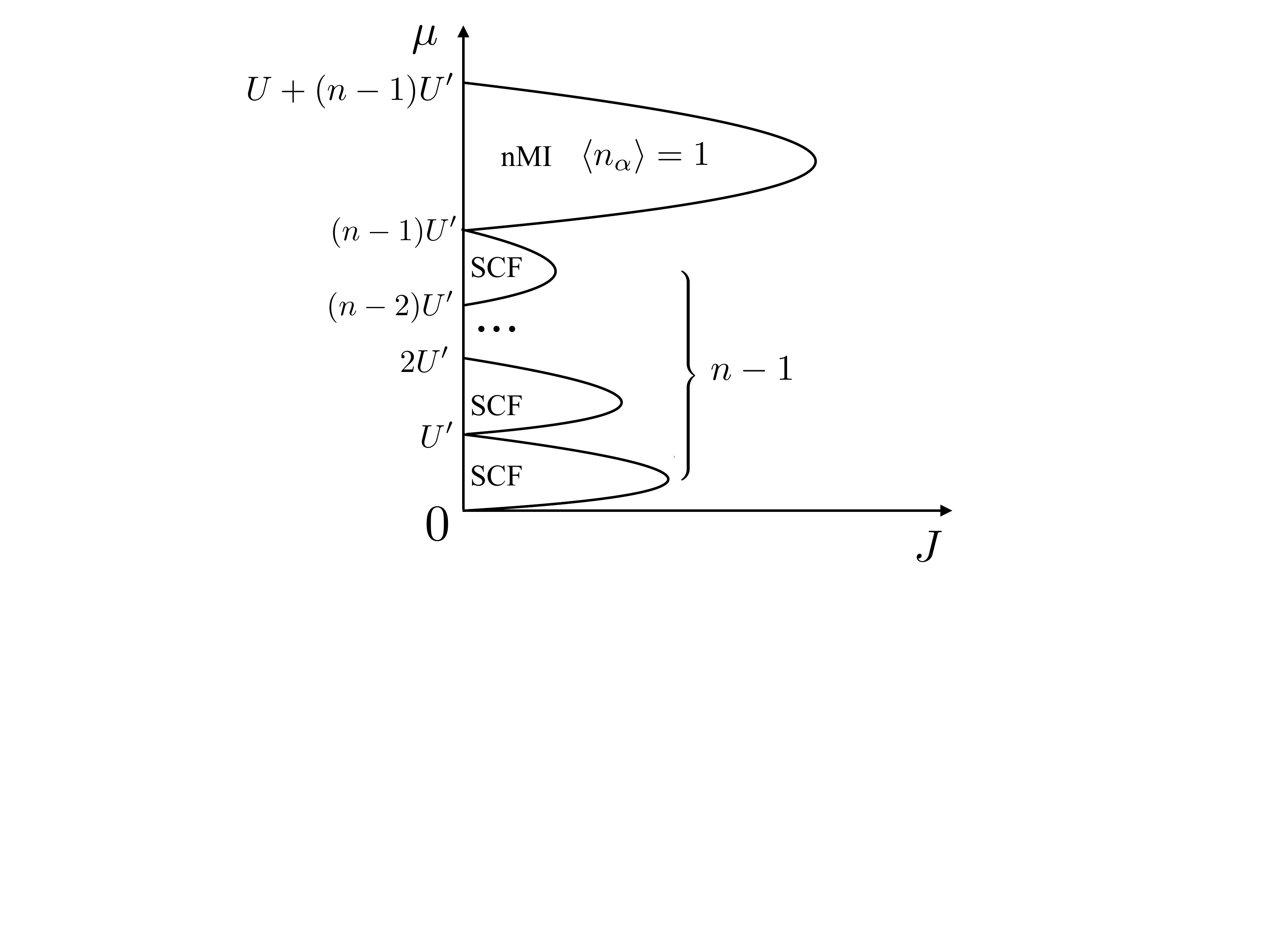}
\caption{\label{Fig5} A sketch phase diagram of the $n$-component ($n\ge2$) bosonic mixture. Here, we only show the lowest $n-1$ SCF lobes and the first nMI lobe.  }
\end{figure}

Using the above expressions, we can obtain the phase boundaries at $J=0$. It is known that there are only nMI and SCF phases at this point, their phase boundaries can be determined by comparing the energies. For the $m$-th nMI lobe, its upper phase boundary should satisfy 
\begin{align}
E_{\mathrm{nMI}}^{m}=E_{\mathrm{SCF-(1)}}^{m+1},
\end{align}
where $E_{\mathrm{SCF-(1)}}^{m+1}$ is the energy of the $(m+1,1)$-th SCF lobe and it can be read as
\begin{align}
E_{\mathrm{SCF-(1)}}^{m+1}=&\frac{U}{2}nm(m-1)+Um+\frac{n(n-1)}{2}U^\prime m^2 \nonumber\\
&+U^\prime(n-1)m-\mu nm-\mu \nonumber\\
=&E_{\mathrm{nMI}}^{m}+Um+U^\prime(n-1)m-\mu.
\end{align}
The upper phase boundary of the $m$-th nMI lobe is then given as 
\begin{align}\label{upb}
\mu_{c,m}^u=Um+U^\prime(n-1)m.
\end{align}
On the other side, the down phase boundary of the $m$-th nMI lobe should satisfy 
\begin{align}
E_{\mathrm{nMI}}^{m}=E_{\mathrm{SCF-(n-1)}}^{m},
\end{align}
where $E_{\mathrm{SCF-(n-1)}}^{m}$ is the energy of the $(m,n-1)$-th SCF lobe. 
Solving this equation, we can get the down phase boundary of the $m$-th nMI lobe,
\begin{align} \label{dpb}
\mu_{c,m}^d=U(m-1)+U^\prime(n-1)m.
\end{align}
From Eq.(\ref{upb}) and Eq.(\ref{dpb}), we can know that the chemical potential width for the $m$-th Mott lobe is $\mu_{c,m}^u-\mu_{c,m}^d=U$ at $J=0$, and this value holds for any interexchange symmetric multi-component mixtures. In addition, we can find that the chemical potential distance from $(m-1)$-th nMI lobe to $m$-th nMI lobe is $\mu_{c,m}^d-\mu_{c,m-1}^u=U^\prime(n-1)$, which indicates that the chemical potential width for each SCF lobe is $U^\prime$ at $J=0$. Based on these results, we can plot a sketch phase diagram for a $n$-component mixture in Fig.~\ref{Fig5}. 

Now, let us study the interspecies entanglement properties of this $n$-component mixture. The interspecies EE for an SF phase is difficult to calculate analytically due to its single-site ground state being varied with the parameters. However, for the SCF and nMI phases, the interspecies EE can be determined because their ground state is known. To do this, we divide the $n$-component mixture into two parts, one is the single component bosons, e.g. boson $a$, and the other one is the remaining $n-1$ components. We denote this partition as $(a|b,c,\cdots)$. We recall that since the interexchange symmetry is preserved, the following partitions are equivalence: $(a|b,c,\cdots)\leftrightarrow(b|a,c,\cdots)\leftrightarrow\cdots$. 

For an nMI state, It is obvious that the interspecies EE is zero. But for the SCF phase, things are more interesting. Considering the ground state of the $(m,k)$-th SCF lobe, the coefficients can be reshaped to a $2\times C_n^k$ matrix after the partition,
\begin{equation}
c=\left(\begin{array}{ccc}
\overbrace{\frac{1}{\sqrt{C_n^k}} \cdots \frac{1}{\sqrt{C_n^k}}}^{C_{n-1}^k} & 0 \cdots 0\\
0  \cdots 0 &  \underbrace{ \frac{1}{\sqrt{C_n^k}} \cdots \frac{1}{\sqrt{C_n^k}}}_{C_{n-1}^{k-1}} 
\end{array}\right)_{2\times C_n^k},
\end{equation}
where the number of rows 2 comes from the fact that there are only two possible Fock states for a single component within an SCF lobe, and they are $|m\rangle$ or $|m-1\rangle$. After tracing out the degrees of freedom of the $n-1$ components, we get the reduced DM for a single component,
\begin{equation} \label{rho_alpha}
\rho_\alpha=cc^\dag=\left(\begin{array}{ccc}
(n-k)/n &  0\\
0   &   k/n
\end{array}\right)_{2\times 2}.
\end{equation}
Hence, we obtain the interspecies EE of the $(m,k)$-th SCF lobe, 
\begin{equation}\label{S}
S=\ln\left[\frac{n}{n-k}\left(\frac{n-k}{k}\right)^{\frac{k}{n}}\right].
\end{equation}
Interestingly, we find that it only depends on the $n$ and $k$, which means $S$ is a constant within an SCF lobe.   

We can give some examples of $S$ here. For a two-component mixture, $n=2$, $k=1$, and $S=\ln2$. For a three-component mixture, $n=3$, $k=1$ or 2, $S$ is the same for the $(m,1)$-th and $(m,2)$-th SCF lobes, and it is equal to $\ln (3 / \sqrt[3]{4})$. All these results are consistent with our numerical calculations in \ref{2CBM} and \ref{3CBM}. For a four-component mixture, $n=4$ and $k=1,2,3$. The interspecies EE is $S^{k=1}=S^{k=3}=\ln(4/3^{3/4})$ and $S^{k=2}=\ln2$. In general, from Eq.~(\ref{rho_alpha}) and Eq.~(\ref{S}) we can know that if $n$ is an even number, then the maximum value of $S$ is $\ln 2$, which corresponds to $k=n/2$.  While if $n$ is an odd number, then the maximum value is $\ln[2n/(n+1)]+\frac{n-1}{2n}\ln[(n+1)/(n-1)]$, which is related to the $(m, (n\pm)/2)$-th SCF lobe. The physical reason is that there exists a maximum number of ways the bosons can be distributed among the $n$ components in these special SCF lobes. For instance, if $n$ is even, the maximum value of $S$ is found in the ($m$, $n/2$)-th SCF lobe. In this lobe, there are $C_n^{n/2}$ degenerated ground states and they are 
\begin{equation}
\left\{
\begin{aligned}
| \underbrace{m-1, \cdots, m-1}_{n/2}, \underbrace{m,\cdots,m}_{n/2}\rangle\\
\cdots\\
| \underbrace{m, \cdots, m}_{n/2}, \underbrace{m-1,\cdots,m-1}_{n/2}\rangle\\
\end{aligned} 
\right.
\end{equation}
The true single-site ground state of this SCF lobe is thus a symmetric summation of the above $C_n^{n/2}$ states. Each of these degenerated states represents a way to distribute the bosons among the $n$ components.  It is obvious that when $k=n/2$ ($n$ is even), the maximum number of ways to distribute the bosons is reached, resulting in a maximum value of $S$. Note that $C_n^{n/2}$ is the ground state degeneracy of this SCF lobe. As a consequence, the maximum value of $S$ corresponds to the maximum ground state degeneracy.  We remind here that all these discussions are in the context of the SCF phase and are based on the single-site ground state. In fact, the relationship between the entanglement entropy and the ground state degeneracy is not always straightforward and can depend on the specific details of the system.  

When $U^\prime$ is sufficiently large, we expect that the demixing effect will occur, indicating that the $n$ component species can not coexist on the same site. Here, we also consider a zero hopping $J\rightarrow 0$ case to obtain a qualitative understanding of the transition point of the demixing effect for this $n$-component mixture.  Similar to the discussion in \ref{3CBM}, we can generalize the relations Eq.~(\ref{conditions}) to a $n$-component mixture situation, 
\begin{equation}\label{c1}
\left\{
\begin{aligned}
E_{nm,0,\cdots,0}&<E_{\mathrm{nMI}}^{m}\\
E_{nm-n+k,0,\cdots,0}&<E_{\mathrm{SCF}-(k)}^m\\
\end{aligned} 
\right.
\end{equation}
Solving the Eq.~(\ref{c1}) at $J=0$ for $k=1,2,3,\cdots, n-1$, we can get the desired results $U^\prime>U$ which is consistent with the calculations in $n=2$ and $n=3$ cases. \\

\section{Conclusions}\label{Conclusions}
In summary, we numerically calculated the mean-field phase diagrams, orders, interspecies EE, and spectral decompositions for a two-  and three-component bosonic mixture under the interexchange symmetry by using the SSGA. Moreover, we extended our discussions to a $n$-component mixture. Interestingly, we find that there are $n-1$ SCF lobes between two neighboring Mott lobes in the phase diagram. We also demonstrated that the chemical potential width for the SCF and nMI lobes at $J=0$ is $U^\prime$ and $U$, respectively, and this result applies to a mixture with any number of components.  Most importantly, we derived the interspecies entanglement entropy for the SCF lobes analytically, and it only depends on the number of components $n$ and the sort number of the SCF lobe $k$. In addition, we studied the mixing effect with a varied interspecies interaction and the mixing-demixing phase transition point $U^c=U^\prime$ is independent of $n$. 

\section{ACKNOWLEDGEMENTS}
This research is supported by the Program of National Natural Science Foundation of China under Grant Nos. 22273067, Scientific Research Fund of Zhejiang Provincial Education Department under Grant No. Y202248878, the Ph.D. research Startup Foundation of Wenzhou University under Grant No. KZ214001P05, and the open project of the state key laboratory of surface physics in Fudan University under Grant No. KF2022$\_$06.

\appendix
\section{The single-site Gutzwiller approach}
The numerical method we used in the main text is the so-called single-site Gutzwiller approach \cite{SingleGutz7} which is suitable for a multi-component bosonic mixture because the dimension of the local Hilbert space is exponential growth with the number of boson components. Let us talk about it in more detail here. As we discussed in the main text,  the wave function of the whole system can be written as $|\Psi\rangle= | i \rangle |\psi\rangle$ in this method, where $|i\rangle$ is the single-site Gutzwiller trial wave function in the basis of local Fock states $|m_a,m_b,m_c \cdots \rangle$ and $|\psi\rangle$ is the wave function of all sites except for site $i$. The exact lattice Hamiltonian can be written as, 
\begin{align}
H=H_{\psi}+H_{i}+H_{\psi i},
\end{align}
where $H_{\psi}$ and $H_{i}$ act only on the respective subsystems. The $H_{\psi i}$ term represents the coupling between $|\psi\rangle$ and $|i\rangle$. We assume that we have known the $|\psi \rangle$, then the Hamiltonian matrix of the whole system in the local Fock state basis $\{|n\rangle=|m_a,m_b,m_c\cdots\rangle\}$ is given by,
\begin{equation}
\begin{aligned}
H_{nn^\prime}&=\langle \psi |H_{\psi} |\psi\rangle \delta_{nn^\prime}+\langle n|H_i|n^\prime \rangle \\ &+\langle \psi | \langle n | H_{\psi i} | n^\prime \rangle |\psi \rangle,
\end{aligned}
\end{equation}
where the first term is a constant energy offset. The wave function on site $i$ can now be obtained by diagonalizing $H_{nn^\prime}$. For our multi-component bosonic mixture Hamiltonian,  only the hopping term is the coupling part, that is,
\begin{equation}
\begin{aligned}
\langle \psi |\langle n | H_{\psi i} |n^\prime \rangle |\psi \rangle  &=-J\langle n | \sum_{\alpha}\left( \alpha_i ^\dag \sum_{\langle j \rangle} \langle \psi| \alpha_j|\psi\rangle \right. \\& \left. +\alpha_i \sum_{\langle j \rangle} \langle \psi| \alpha_j^\dag|\psi\rangle \right) |n^\prime \rangle,
\end{aligned}
\end{equation}
where $\alpha_i$ is the $\alpha$-component boson annihilation operator on site $i$ and $\langle j \rangle$ indicates the summation over all nearest neighbors of site $i$. It clearly shows that in the single-site Gutzwiller method, the multi-component Bose-Hubbard model is treated as a single lattice site couples only to the average mean field $\langle \psi| \alpha_j|\psi\rangle $ and its conjugate. Thus, we get the mean-field Hamiltonian Eq.~(\ref{HamSSGA}). By diagonalizing $H_{nn^\prime}$, a new expectation value of the superfluid order parameter $\langle i| \alpha_i|i\rangle $ for the site $i$ can be computed. Using a self-consistent loop, the Hamiltonian can be solved without initial knowledge of $|\psi\rangle$. This algorithm can be described as the following steps: 
\begin{itemize}
\item Step 1: initializing a random value $\langle \psi| \alpha_j|\psi\rangle$;
\item Step 2: diagonalizing $H_{nn^\prime}$;
\item Step 3: calculating $\langle i| \alpha_i|i\rangle $;
\item Step 4: replace $\langle \psi| \alpha_j|\psi\rangle$ with $\langle i| \alpha_i|i\rangle $;
\item Step 5: repeat steps 2-4 until it converges.  
\end{itemize}

Usually, the single-site wave function $|i\rangle$ is determined by imaginary time evolution \cite{SingleGutz5}. However, the above self-consistent diagonalization scheme can be used to determine it as well \cite{SingleGutz7}. In actual calculations, there are some tricks that need to be carefully dealt with. Firstly, the single-site ground state of the SCF phase is degenerate, and thus all measurements should be calculated using the true ground state, which is a symmetric sum of all degenerate states. Secondly, low-accuracy convergence can lead to unphysical noise in the measurement of the SCF order parameter. These noises can be removed by setting a very high convergence precision.

\section{The apparently random behavior of the measurements when the demixing effect occurs}
\begin{figure}[htbp] 
\centering
\includegraphics[width=0.45\textwidth]{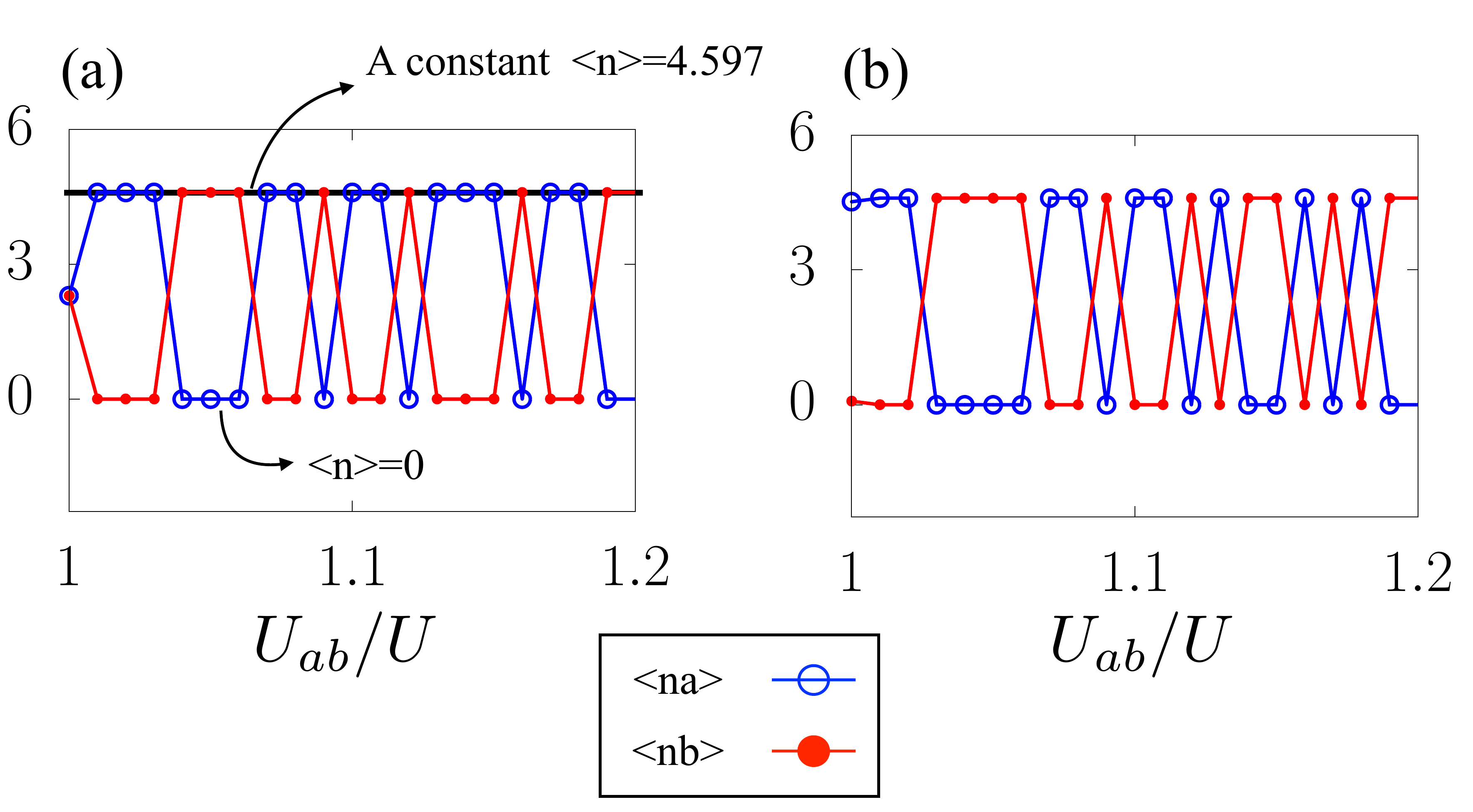}
\caption{ \label{FigB1}(Color online) Details of Fig.~\ref{Fig2} (a) at $U_{ab}/U>1$. It clearly shows that if $\langle n_a \rangle$ ($\langle n_b \rangle$) is non-zero, then $\langle n_b \rangle$ ($\langle n_a \rangle$) must be zero, implying that the mixture is in a demixed state.  (a) and (b) represent the outcomes of running the self-consistent process twice. }
\end{figure}
As we can see in Fig.~\ref{Fig2} (a)(b) and Fig.~\ref{Fig4} (a)(b), the locations of sudden changes at $U_{ab}/U > 1$ appear random. These locations have no special physical meanings because they depend on the initial conditions of the numerical method. As an example, we plot Fig.~\ref{Fig2} (a) by running the self-consistent process twice and presenting it in Fig.~\ref{FigB1}. It shows that the location of the sudden changes can not be reproduced. However, we find that the value $\langle n \rangle=4.597$ is a constant that is independent of the self-consistent procedure. This special value is only dependent on the parameters of the Hamiltonian which makes it meaningful.

%\bibliography{manuscript}
\input{manuscript.bbl}
\end{document}